\documentclass[preprint,aps]{revtex4}
\usepackage{graphicx}
\usepackage{dcolumn}
\usepackage{bm}
\usepackage{amssymb}
\usepackage{amsmath}
\newtheorem{lemma}{Lemma}
 \newenvironment{Proof.}[1][Proof.]{\begin{trivlist}
     \item[\hskip \labelsep {\bfseries #1}]}{\end{trivlist}}
\begin{document}
\setcounter{page}{1}
\title
{Maximal extension of the Schwarzschild spacetime inspired by
noncommutative geometry}
\author
{I. Arraut$^1$, D. Batic$^{2,}$$^3$ and M. Nowakowski$^1$}
\affiliation{$^1$
Departamento de Fisica, Universidad de los Andes,
Cra.1E No.18A-10, Bogota, Colombia\\
$^2$ Departamento de Matematicas, Universidad de los Andes,
Cra 1E, No. 18A-10, Bogota, Colombia\\
$^3$
Department of Mathematics,\\
University of West Indies at Mona, Kingston 6, Jamaica }







\begin{abstract}
We derive a transformation of the noncommutative geometry inspired Schwarzschild
solution into new coordinates such that the apparent
unphysical singularities of the metric are removed.
Moreover, we give the maximal singularity-free atlas for the
manifold with the metric under consideration. This atlas
reveals many new features e.g. it
turns out to describe an infinite
lattice of asymptotically flat universes connected by black hole tunnels.\end{abstract}


\maketitle

\section{Introduction}
The area of quantum gravity has not yet converged into a single
theory and at present several rival theories co-exist.
Nevertheless, certain common or global features like
noncommutativity at lengths less than $10^{-16}cm$ \cite{Nico1,
Rest1}, a new uncertainty principle including gravity effects
\cite{Adler1}, the avoidance of physical singularities
\cite{Avoidence} (e.g. replaced in the noncommutative geometry by
a deSitter core), black hole remnants \cite{Remnants} etc. are
expected. The noncommutative aspect of spacetime has been recently
applied to the final state of a black hole \cite{Nico1, Nicoex1,
Nicolini, Nico2, Nico3, Nicoex2}. The starting point of these new
developments is  the commutation relation $[x^{\mu},
x^{\nu}]=\theta^{\mu \nu}$. Based on such a commutation relation
one can show that one of the effect of non commutativity is to
replace point-like objects by de-localized matter sources which
turn to be of Gaussian form. Following \cite{Spal1, Nicoex2,
Nicoex3} we can take, instead of the point mass $M$, described by
a $\delta$-function distribution, a static, spherically symmetric,
Gaussian-smeared matter source
\[
\rho =\frac{M}{(4\pi\theta)^{D/2}}e^{-\frac{r^2}{4\theta}},
\]
where $D>0$ is the dimension of the underlying manifold. This
observation gave rise to new models of mini black holes
\cite{Nico1, Nicoex1, Nicolini, Nico2, Nico3, Nicoex2,we} where
the singularity at the origin is replaced by a self-gravitating
droplet. Although the issue of smearing point-like structures
might not be the only fingerprint of the noncommutative geometry,
these models explicitly reveal its importance. For instance, the
central singularity is replaced by a deSitter core (droplet) and
the metric can have two horizons depending whether the black hole
mass exceeds a certain critical mass \cite{Nicolini}. In
Schwarzschild coordinates, these horizons leave unphysical
singularities in the metric components. We will present a
coordinates extension in the following sections of the paper. It
is therefore of some importance to find a maximal atlas for this
metric and its interpretation. For any metric in General
Relativity with apparent unphysical singularities there is a
continued interest in finding maximal singularity-free extensions
\cite{Kruskal, Grav, Rest2}. Such maximal atlases often shed new
light on the manifold under consideration. In particular, this is
true for metrics which are partly motivated by a quantum
mechanical property such as the noncommutativity of the coordinate
operators. New phenomena closely related to this quantum nature of
spacetime can emerge. Indeed, for the metric inspired by
noncommutative geometry which we study in the present paper, the
maximal atlas reveals the existence of black hole tunnels
connecting parallel universes.

The paper is organized as follows. In section II we will discuss
the singularities of the metric which emerge in the framework of
noncommutative geometry when a point-like structure is smeared by
the so-called Gaussian prescription. Section III treats the new
transformation of the Schwarzschild coordinates which leads to the
maximal atlas. Section IV continues these considerations and
discusses the determination of the constants in the
transformation. Section V interprets the results in terms of a
Penrose diagram. Section VI is devoted to the extreme case where
the mass is equal to a critical mass. In section VII we draw our
conclusions.
\section{Singularities of the metric}
The replacement of the sharp point-like structure discussed above suggests that the metric can be based on the Gaussian
mass distribution,
\begin{equation}\label{density}
\rho_{\theta}(r)=\frac{M}{(4\pi\theta)^{3/2}}~e^{-r^2/(4\theta)}.
\end{equation}
The ansatz of an anisotropic perfect fluid energy-momentum tensor taken together with two
equations of state in which the pressure is determined by the
Tolman-Oppenheimer-
Volkov equation leads to
the noncommutative geometry inspired Schwarzschild solution \cite{Nicolini} given by
\begin{equation}\label{NCSS}
ds^2=\left(1-\frac{4M}{\sqrt{\pi}r}\gamma\left(\frac{3}{2},\frac{r^2}{4\theta}\right)\right)dt^2
-\left(1-\frac{4M}{\sqrt{\pi}r}\gamma\left(\frac{3}{2},\frac{r^2}{4\theta}\right)\right)^{-1}dr^2
-r^2d\vartheta^2-r^2\sin^2{\vartheta}~d\varphi^2,
\end{equation}
where $M$ is the mass of the black hole, $\theta>0$ is a parameter encoding noncommutativity and $\gamma$
is the incomplete lower gamma function. The singularities of the metric are determined by the equation
\begin{equation}\label{zeroes}
g_{00}(r):=1-\frac{4M}{\sqrt{\pi}r}\gamma\left(\frac{3}{2},\frac{r^2}{4\theta}\right)=0.
\end{equation}
According to \cite{Nicolini} there are three possible scenarios, namely
\begin{enumerate}
\item
if $M<M_0\approx 1.9\sqrt{\theta}$, the function $g_{00}$ never vanishes,
\item
if $M=M_0$, $g_{00}$ vanishes just at one value $r_0\approx 3.0\sqrt{\theta}$ (extremal black hole),
\item
if $M>M_0$, there exist $r_\pm$ with $0<r_{-}<r_{+}$ such that $g_{00}(r\pm)=0$
\end{enumerate}
A striking feature of the metric (\ref{NCSS}) is the absence of a true singularity at $r=0$.
Moreover, the above classification is of numerical nature since the equation $g_{00}(r)=0$
cannot be solved in closed form. We prove below
that in the case that the black hole mass exceeds the critical mass $M_0$ the roots
of the equation $g_{00}(r)=0$ are simple. This result will play an important
role in the following considerations. First we need to clarify the nature of the critical mass $M_0$,
which turns out to be the minimal mass to have horizons. Therefore, if we employ the horizon
equation to define a function $M=M(r_H)$, where $r_H$ is a solution of the equation $g_{00}(r)=0$, we obtain
\begin{equation}
M(r_H)\equiv \frac{\sqrt{\pi}}{4}\frac{r_H}{\gamma\left(3/2; r_h^2/4\theta\right)}.
\end{equation}
Thus, having considered the derivative $dM(r_H)/dr_H$, we can look for $r_0$ such that the latter vanishes, i.e.
$\left. dM/dr_H\right|_{r_0}=0$. We do this in order to define $M_0\equiv M(r_0)$. One can easily
show that the above derivative vanishes if and only if
\begin{equation}
\gamma\left(3/2; r_H^2/4\theta\right)=r_H\frac{d\gamma}{dr_H}.
\label{extreme}
\end{equation}
From the properties of the gamma function one can easily derive the following result
\begin{equation}
\gamma\left(3/2; r_H^2/4\theta\right)=\frac{1}{4\theta^{3/2}}\int_0^{r_H}\, dt\, t^2 e^{-t^2/4\theta}
\end{equation}
and as a consequence
\begin{equation}
\gamma\left(3/2; r_H^2/4\theta\right)=\frac{1}{4\theta^{3/2}}\, r_H\, r_m^2 e^{-r_m^2/4\theta}
\end{equation}
where $r_m\in[\,0,\,r_H]$. Eq. (\ref{extreme}) can now be written as
\begin{equation}
\frac{1}{4\theta^{3/2}}\, r_H\, r_m^2 e^{-r_m^2/4\theta}=\frac{1}{4\theta^{3/2}}\,r_H^3 e^{-r_H^2/4\theta},
\end{equation}
which admits a unique solution if and only if $r_H=r_m=r_0$. We can conclude that there exists
a unique horizon radius $r_0$, corresponding to the critical mass $M_0$. With this in mind
we can establish the following lemma
\begin{lemma}\label{lemma1}
Let $M>M_0$ and $0<r_{-}<r_{+}$ such that $g_{00}(r_\pm)=0$. Then, $r_{-}$ and $r_{+}$
are simple zeroes of (\ref{zeroes}).
\end{lemma}
\begin{Proof.}
We give the proof for a generic solution $r_{H}$ of the horizon equation $g_{00}(r_H)=0$.
Therefore $r_H$ corresponds either to $r_+$ or $r_-$. Notice that if the limit
\[
\lim_{r\to r_{H}}\frac{g_{00}(r)}{r-r_{H}}
\]
is finite, then $r_{H}$ is a simple root. Since $g_{00}$ is differentiable on the interval
$[0,\infty)$ we can expand it in a Taylor series and we obtain
\[
g_{00}(r)=(r-r_{H})\left[g^{'}_{00}(r_{H})+\mathcal{O}(r-r_{H})\right].
\]
Hence,
\[
\lim_{r\to r_{H}}\frac{g_{00}(r)}{r-r_{H}}=g^{'}_{00}(r_{H})
\]
and in order to show that $r_{H}$ is a simple zero we need to prove that $g^{'}_{00}(r_{H})\neq 0$.
Taking into account the fact that
\[
g^{'}_{00}(r_H)=\frac{4M}{\sqrt{\pi}r_H}\left[\frac{1}{r_H}\gamma\left(\frac{3}{2},
\frac{r_H^2}{4\theta}\right)-\gamma^{'}\left(\frac{3}{2},\frac{r_H^2}{4\theta}\right)\right]
\]
where a prime denotes differentiation with respect to the horizon
radius and comparing the above equation with the Eq.
(\ref{extreme}), we can see that $g^{'}_{00}(r_H)$ vanishes if and
only if $r_H=r_0$. This implies that $M=M_0$ which is at variance
with the initial assumption. As a result we can conclude that
$g_{00}^{'}(r_{H})\neq 0$ for $M>M_0$ and $r_H\neq r_0$.
\end{Proof.}

\section{A new transformation}
We show that the singularities of (\ref{NCSS}) can be removed by a suitable
coordinate transformation as in the case of the Reissner-Nordstr\"om solution. In order to do that
we shall follow \cite{Grav}. Like in the Kruskal approach \cite{Kruskal} we introduce coordinates
$u(t,r)$ and $v(t,r)$ such that the original metric goes over to
\begin{equation}\label{transf}
ds^2=f^2(u,v)(dv^2-du^2)-r^2(u,v)(d\vartheta^2+\sin^2{\vartheta}~d\varphi^2)
\end{equation}
with the requirement that $f^2\neq 0$. This will happen if $u$ and $v$ satisfy the
non homogeneous system of first order nonlinear partial differential equations
\begin{eqnarray}
&&f^2(u,v)\left[(\partial_t v)^2-(\partial_t u)^2\right]=g_{00}(r),\label{tre}\\
&&f^2(u,v)\left[(\partial_r v)^2-(\partial_r u)^2\right]=-g_{00}^{-1}(r),\label{quattro}\\
&&\partial_r u~\partial_t u-\partial_r v~\partial_t v=0. \label{cinque}
\end{eqnarray}
The next step is to find a suitable transformation of the variable $r$ such that the above system becomes a
homogeneous system of PDEs. If we multiply (\ref{quattro}) by $g_{00}^2$
and we introduce a new spatial variable $r_{*}=r_{*}(r)$ defined through
\begin{equation}\label{stella}
\frac{dr_{*}}{dr}=\frac{1}{g_{00}(r)}
\end{equation}
then we have
\begin{eqnarray}
&&(\partial_t v)^2-(\partial_t u)^2=F(r_{*}),\label{Kru_4}\\
&&(\partial_{r_{*}} v)^2-(\partial_{r_{*}} u)^2=-F(r_*),\label{Kru_5}\\
&&\partial_t v~\partial_{r_{*}} v-\partial_t u~\partial_{r_{*}} u=0\label{Kru_6}
\end{eqnarray}
with $u=u(t,r_*)$, $v=v(t,r_*)$ and $F(r_{*}):=g_{00}(r)/f^2$ where $r$ is a function of $r^{*}$.
We want to show that $u$ and $v$
satisfy a wave equation. If we consider the combinations (\ref{Kru_4})$+$(\ref{Kru_5})$\pm 2$(\ref{Kru_6})
we arrive at the following
equations
\begin{eqnarray}
&&(\partial_t v+\partial_{r_{*}} v)^2=(\partial_t u+\partial_{r_{*}} u)^2,\label{Kru_7}\\
&&(\partial_t v-\partial_{r_{*}} v)^2=(\partial_t u-\partial_{r_{*}} u)^2. \label{Kru_8}
\end{eqnarray}
While taking the square roots of the above equations only those choices
of the sign are allowed for which the determinant of the Jacobian
of the transformation $\widetilde{x}:=(v,u,\vartheta,\varphi)\longrightarrow x=(t,r,\vartheta,\varphi)$
does not vanish identically. Hence, we require
\begin{equation}\label{Kru_9}
\mbox{det}(J)=\mbox{det}\left(\begin{array}{cc}
\partial_t v & \partial_{r_{*}}v\\
\partial_t u & \partial_{r_{*}}u
\end{array}\right)=\partial_t v~\partial_{r_{*}}u-\partial_{r_{*}}v~\partial_t u\neq 0.
\end{equation}
Clearly, there are four possible choices of the sign. If we consider
for example the case
\[
\partial_t u+\partial_{r_{*}} u=\partial_t v+\partial_{r_{*}} v,
\quad \partial_t u-\partial_{r_{*}} u=\partial_t v-\partial_{r_{*}} v,
\]
we would obtain $\partial_t u=\partial_t v,~\partial_{r_{*}} u=\partial_{r_{*}} v$
which implies $\mbox{det}(J)=0$. The same happens if we consider
\[
\partial_t u+\partial_{r_{*}} u=-\partial_t v-\partial_{r_{*}} v,\quad
\partial_t u-\partial_{r_{*}} u=-\partial_t v+\partial_{r_{*}} v.
\]
On the other hand the choice
\[
\partial_t u+\partial_{r_{*}} u=\partial_t v+\partial_{r_{*}} v,\quad
\partial_t u-\partial_{r_{*}} u=-\partial_t v+\partial_{r_{*}} v,
\]
leads to the equations
\begin{equation}\label{pre_wave}
\partial_t u=\partial_{r_{*}}v,\qquad \partial_{r_{*}} u=\partial_t v.
\end{equation}
We require that
\begin{equation}\label{Kru_10}
\mbox{det}(J)=(\partial_{r_{*}}u)^2-(\partial_t u)^2\neq 0
\end{equation}
in that part of the manifold which is described by the coordinates $(v,u,\vartheta,\varphi)$.
In the next section we will show that the above
condition is indeed satisfied. Finally, it is not difficult to verify that the choice
\[
\partial_t u+\partial_{r_{*}} u=-\partial_t v-\partial_{r_{*}} v,\quad
\partial_t u-\partial_{r_{*}} u=\partial_t v-\partial_{r_{*}} v
\]
is equivalent to the previous one in the sense that both give rise to the same wave equations.
Thus, from (\ref{pre_wave}) we can
derive the following wave equations:
\[
\partial_{tt}u-\partial_{r_{*}r_{*}}u=0,\qquad \partial_{tt}v-\partial_{r_{*}r_{*}}v=0,
\]
with the solutions
\begin{equation}\label{sol_wave}
u(t,r_{*})=h(r_{*}+t)+g(r_{*}-t),\qquad
v(t,r_{*})=h(r_{*}+t)-g(r_{*}-t).
\end{equation}
Substituting (\ref{sol_wave}) into (\ref{Kru_4}) or (\ref{Kru_5})
gives
\begin{equation}\label{condition}
4\frac{dh}{dy}\frac{dg}{dz}=F(r_{*})
\end{equation}
with $y:=r_{*}+t$, $z:=r_{*}-t$ whereas (\ref{Kru_6}) gives a
trivial identity. Note that (\ref{condition}) fixes the relative signs of the functions
$h$ and $g$ since by definition of the function $F(r_{*})$ we have
$F(r_{*})>0$ for $r>r_{+}$ and  $F(r_{*})<0$ for $r_{-}<r<r_{+}$.
Moreover, if we substitute (\ref{sol_wave}) into (\ref{Kru_10}) the
invertibility condition simplifies to the requirement
\begin{equation}\label{Kru_11}
F(r_{*})\neq 0.
\end{equation}
Clearly, (\ref{Kru_11}) is not satisfied for $r=r_{\pm}$. This means
that on the spheres with radius $r_{\pm}$ the transformations from
spherical coordinates to ones which we are constructing, are not invertible. However, this is not
really a problem since our goal is to construct several charts patching different regions of the manifold
and by construction we will see that the transfer functions between these charts are always invertible.
Finally, if we compute
$\partial_{r_{*}}(\ref{condition})/(\ref{condition})$ and
$\partial_{t}(\ref{condition})/(\ref{condition})$ with the
requirement that $r\neq r_{\pm}$ we end up with the following equations:
\[
\left(\frac{d^2 h}{dy^2}\right)/\left(\frac{dh}{dy}\right)+\left(\frac{d^2
g}{dz^2}\right)/\left(\frac{dg}{dz}\right)=\frac{1}{F}\frac{dF}{dr_{*}},\qquad
\left(\frac{d^2 h}{dy^2}\right)/\left(\frac{dh}{dy}\right)-\left(\frac{d^2
g}{dz^2}\right)/\left(\frac{dg}{dz}\right)=0.
\]
Summing the above equations we obtain
\begin{equation}\label{eq_1}
2\frac{d}{dy}\left(\ln{\frac{dh}{dy}}\right)=\frac{d}{dr_{*}}\left(\ln{F}\right).
\end{equation}
Hence, once we solve for $h$ the unknown function $g$ can be
determined from
\begin{equation}\label{eq_2}
\frac{d}{dz}\left(\ln{\frac{dg}{dz}}\right)=\frac{d}{dy}\left(\ln{\frac{dh}{dy}}\right).
\end{equation}
Since the variables $y$ and $r_{*}$ in (\ref{eq_1}) can be regarded
as independent variables, we can set both sides in (\ref{eq_1}) equal
to a separation constant $2\gamma$. The factor $2$ has been introduced
in order to simplify the left hand side of (\ref{eq_1}). Hence,
the solutions read
\[
F(r_{*})=c_{1}e^{2\gamma r_{*}},\qquad h(y)=\frac{c_2}{\gamma}~e^{\gamma
y}+c_3,\quad g(z)=\frac{c_4}{\gamma}~e^{\gamma z}+c_5,
\]
with $\gamma\neq 0$. However, this condition on $\gamma$ is always
satisfied as we shall see in the next section. Guided by the principle
that we wish to derive the most simple expressions for $u$ and $v$ we can choose without loss of generality,
$c_3=c_5=0$. Taking into account the definition of the tortoise
coordinate $r_{*}$ we obtain,
\[
v(t,r_{*})=\frac{e^{\gamma r_{*}}}{\gamma}\left(c_2~e^{\gamma t}-c_4~e^{-\gamma t}\right),\quad
u(t,r_{*})=\frac{e^{\gamma r_{*}}}{\gamma}\left(c_2~e^{\gamma t}+c_4~e^{-\gamma t}\right).
\]
\subsubsection{Regions I and III ($r>r_{+}$)}
Let $(v_I,u_I)$ denote the specific coordinates $(v,u)$ specialized for the
region $I$ characterized by $r>r_{+}$. In this region $F(r_{*})>0$ and
equation (\ref{condition}) requires that we choose $h$ and $g$ with a positive relative sign.
This implies that the constants $c_2$ and $c_4$
have to be chosen with the same sign. The simplest choice is
$c_{2}=\gamma/2=c_4$. Thus, we have
\[
v_{I}(t,r_{*})=e^{\gamma r_{*}}\sinh{(\gamma t)},\quad u_{I}(t,r_{*})=e^{\gamma r_{*}}\cosh{(\gamma t)}
\]
and
\[
f^{2}=\frac{g_{00}(r)}{c_1}~e^{-2\gamma r_{*}}.
\]
In the next section we shall see that the requirement $f^{2}\neq 0$ fixes the value of the separation
constant $\gamma$. The constant $c_1$ can be fixed by requiring that the inverse
transformation from the coordinates $(v_I,u_I,\vartheta,\varphi)$ to
the coordinates $(t,r,\vartheta,\varphi)$ gives (\ref{NCSS}) again and
we find that $c_1=\gamma^2$. Thus, we conclude that
\[
f^{2}=\frac{g_{00}(r)}{\gamma^2}~e^{-2\gamma r_{*}}.
\]
Clearly, we still have the possibility to choose the functions $h$ and $g$
both with negative signs so that their relative sign is
again positive. In this case we get a new set of coordinates representing a
portion of the manifold (we call it region $III$) which is isometric
to region $I$. In particular, the coordinates which map such a region
are
\[
v_{III}(t,r_{*})=-e^{\gamma r_{*}}~\sinh{(\gamma t)},\quad
u_{III}(t,r_{*})=-e^{\gamma r_{*}}~\cosh{(\gamma t)}.
\]
Note that $u_{III}$ is always negative as it should be according
to the present choice of the relative sign of the functions $h$ and
$g$ whereas $v_{III}$ can assume both negative and positive values.
\subsubsection{Regions II and IV ($r_{-}<r<r_{+}$)}
Region $II$ is obtained by deriving the corresponding transformation when the radial coordinate varies in the interval
$(r_{-},r_{+})$. In this case $F(r^{*})<0$ and the relative sign of the
functions $h$ and $g$ must be negative. If we choose $c_4=-\gamma/2=-c_2$ we get
\[
v_{II}(t,r_{*})=e^{\gamma r_{*}}~\cosh{(\gamma~t)},\quad
u_{II}(t,r_{*})=e^{\gamma r_{*}}~\sinh{(\gamma~t)}.
\]
Clearly, we are also free to make the opposite choice. This is equivalent to the transformation
$(v,u)\longrightarrow (-v,-u)$. Thus, region $IV$ is isometric to
region $II$ and it is described by the coordinates
\[
v_{IV}(t,r_{*})=-e^{\gamma r_{*}}~\cosh{(\gamma~t)},\quad
u_{IV}(t,r_{*})=-e^{\gamma r_{*}}~\sinh{(\gamma~t)}.
\]
In the next section we shall discuss that the overlapping conditions at $r=r_{\pm}$ are satisfied.
It is evident that the inverse transformations can only be given implicitly since
from (\ref{stella}) we see that $r_{*}(r)$ cannot be inverted in terms of elementary functions.
However, from the following relations
\begin{equation}\label{undici}
u^2-v^2=e^{2\gamma r_{*}},\quad \frac{1}{\gamma}\tanh^{-1}\left(\frac{v}{u}\right)=t,
\end{equation}
we see that in the $(v,u)$-plane,the lines $t=$const are straight lines $v/u=$const
whereas lines $r=$const are represented by the hyperbolae $u^2-v^2=$const.

\section{Determination of the separation constant $\gamma$}
Let us first consider the singularity at $r_+$. Lemma~\ref{lemma1} ensures
that $r_+$ is a simple zero of the metric coefficient $g_{00}$. Thus, the following
Taylor expansion holds in a neighborhood of $r_+$, namely
\[
g_{00}(r)=\alpha(r-r_+)+\beta(r-r_+)^2+\mathcal{O}(r-r_+)^3
\]
with
\[
\alpha:=g_{00}^{'}(r_+),\quad\beta:=\frac{1}{2}g_{00}^{''}(r_+).
\]
On the other hand the expansions of the tortoise coordinate $r_{*}$ and of the functions $e^{\pm 2\gamma r_{*}}$ are
\begin{eqnarray*}
r_{*}(r)&=&\frac{1}{\alpha}\ln{(r-r_+)}-\frac{\beta}{\alpha^2}(r-r_+)+\mathcal{O}(r-r_+)^2,\\
e^{\pm 2\gamma r_{*}}&=&(r-r_+)^{2\gamma/\alpha}\left[1\mp\frac{2\beta\gamma}{\alpha^2}(r-r_+)
+\mathcal{O}(r-r_+)^2\right].
\end{eqnarray*}
Hence, we have
\[
\frac{g_{00}(r)}{e^{2\gamma r_{*}}}=(r-r_{+})^{(\alpha-2\gamma)/\alpha}\left[\alpha+\left(\beta
+\frac{2\beta\gamma}{\alpha}\right)(r-r_+)^2+\mathcal{O}(r-r_+)^2\right].
\]
From Lemma~\ref{lemma1} it follows that $\alpha$ can never vanish
and we can choose $\gamma$ so that the singularity at $r_+$ is
cancelled. This happens if $\gamma$ is chosen to be
\[
\gamma_+=\frac{\alpha}{2}=\frac{1}{2}g^{'}_{00}(r_+).
\]
Clearly, it is not possible to cancel both singularities at once. In order
to remove the singularity at $r_-$ we can proceed as we did above and we find that
\[
\gamma_{-}=\frac{1}{2}g^{'}_{00}(r_{-}).
\]
If we choose $\gamma=\gamma_{+}$ in the $(v,u)$ coordinates we can
proceed from an arbitrarily large $r$ towards smaller $r$ across
the set $r\in(r_{-},r_{+}$. If we want to continue further across
$r=r_{-}$ we have to go back to the coordinates $(v,u)$ and choose
$\gamma=\gamma_{-}$. In this way we can continue across $r=r_{-}$
and reach $r=0$ since it is not a singularity for the metric
(\ref{NCSS}) and even continue through this timelike surface.
\section{Radial observers}
We now analyze the motion of a radial observer inside the
noncommutative geometry inspired Schwarzschild black hole. In
doing so we will not address issues of stability which we postpone
to a future investigation. When the observer enters the event
horizon at $r_{+}$ but has still not crossed the second horizon at
$r_{-}$, the radial coordinate $r$ becomes timelike, implying that
the motion proceeds with decreasing $r$. Once the observer has
crossed the Cauchy horizon at $r_{-}$, the coordinate $r$ becomes
spacelike again. This means that we have two possible kinds of
motion given by increasing and decreasing values of $r$. At this
point the observer can take two decisions: either to cross the
timelike surface $r=0$ in order to approach an asymptotically flat
region or to reverse his/her course as is the case in the
Reissner-Nordstr\"{o}m metric \cite{Poisson}. By taking the latter
decision, the observer will cross another copy of the surface
$r=r_{-}$. Having entered the new region $r_{-}<r<r_{+}$ the
radial coordinate becomes again timelike and the observer is
forced to cross a new copy of the event horizon $r_{+}$. In this
way the observer will emerge out of a white hole in an
asymptotically flat universe. However, the journey of the observer
might not finish here since he still has the possibility to enter
the noncommutative inspired Schwarzschild black hole living in
this new universe.

\begin{figure}
\begin{center}
\scalebox{1}{\includegraphics{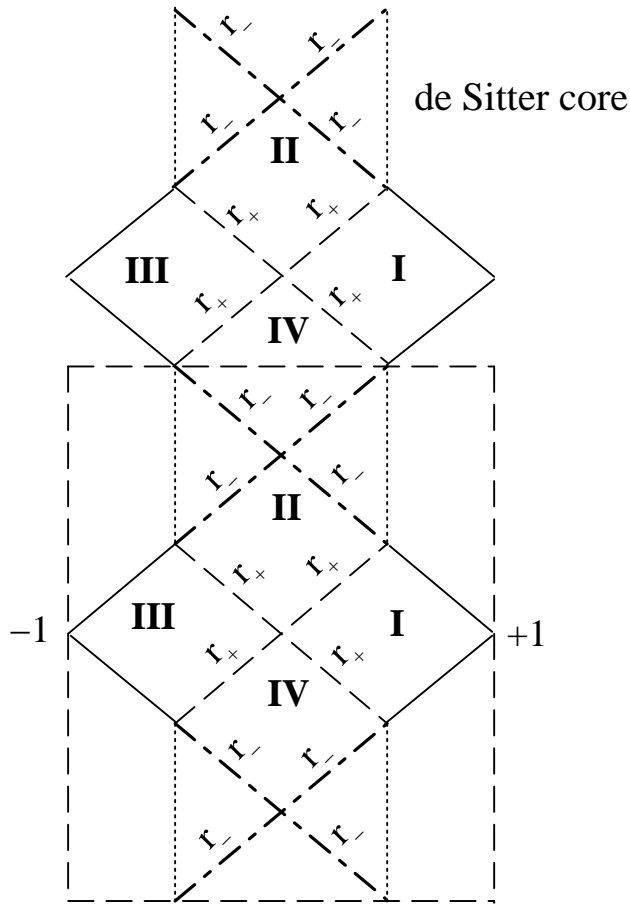}} \caption{The conformal
diagram of the maximally extended noncommutative inspired
Schwarzschild spacetime. $r_{+}$ and $r_{-}$ represent the event
and Cauchy horizons, respectively. The central singularity
appearing in the Reissner-Nordstr\"{o}m metric is now replaced by
a regular deSitter core (dotted line). The upper and lower part of
the box indicated by the dashed line can be identified to make the
manifold cyclic in the time coordinate.}
\end{center}
\end{figure}
\label{myfigure}

The result is the possibility to make a trip through an infinite
number of universes connected by black hole tunnels. In order to
depict the dynamics described above we shall construct a Penrose
diagram (see Fig. 1) for the spacetime structure of the maximal
extended Schwarzschild solution inspired by noncommutative
geometry. First of all, we observe that the radial null geodesics
in the metric (\ref{transf}) are $du/dv=\pm 1$. For this reason it
is convenient to switch to null coordinates $p=u+v$, $q=u-v$ with
$v=e^{\gamma_{+}r_{*}}\sinh{(\gamma_{+}t)}$ and
$u=e^{\gamma_{+}r_{*}}\cosh{(\gamma_{+}t)}$ in order to apply the
so-called Penrose transformation $P=\tanh{p}$ and $Q=\tanh{q}$. It
is clear from (\ref{undici}) that the equation of the apparent
singularity at $r_{+}$ reads $u^2-v^2=0$. Thus, in the
$(Q,P)$-plane the equation $u=v$ becomes $Q=0$ whereas $u=-v$
reads $P=0$. Moreover, the null infinities $p=\pm\infty$ and
$q=\pm\infty$ are mapped into $P=\pm 1$ and $Q=\pm 1$. In this way
we mapped the $(v,u)$-plane in the square $[-1,1]\times[-1,1]$
where as usual we can introduce new coordinates $U=(P+Q)/2$ and
$V=(P-Q)/2$. We recall that the coordinate singularity at $r_{+}$
has equation $U=\pm V$. The null infinities are now straight line
segments with equations $U+V=P=\pm 1$ and $U-V=Q=\pm 1$, while the
lines $r=$const in the subspace
$\{\vartheta=\mbox{const},~\varphi=\mbox{const}\}$ which are
represented by the hyperbolae $u^2-v^2=$const are still hyperbolae
in the $(U,V)$-plane with equation \cite{Ple}
\[
\alpha^2\left[\left(U+\frac{1}{\alpha}\right)^2-V^2\right]=1,\quad \alpha:=\frac{1-C}{1+C},
\quad C:=e^{2pq}=\mbox{const}.
\]
Concerning the regions $r<r_{+}$ we just repeat the above
procedure with $v=e^{\gamma_{-}r_{*}}\sinh{(\gamma_{-}t)}$ and
$u=e^{\gamma_{-}r_{*}}\cosh{(\gamma_{-}t)}$. In this way we can
patch together conformal diagrams of different parts of the
original manifold and we end up with the Penrose diagram shown in
the rectangle in the centre of Fig.1. Radial null geodesics would
be represented by straight lines parallel to the horizons
$r_{\pm}$. Since no future-directed null geodesic can go from
region $II$ to region $I$ or $III$ we conclude that $r=r_{+}$ is
an event horizon. Note that the present situation is very
different from the classical Schwarzschild or the
Reissner-Nordstr\"{o}m case since there is no central singularity
at all. This point allows to interpret the noncommutative inspired
Schwarzschild solution as a series of open tunnels connecting
infinitely many asymptotically flat universes. However, there is
also an alternative interpretation when we restrict our attention
to the thick-line rectangle of Fig.1. In fact, we realize that the
upper tunnel (the strip where in the Reissner-Nordstr\"{o}m case
we would expect to have the central singularity) is a copy of the
lower tunnel. This suggests that we might identify the two
tunnels. If we do that, the manifold becomes finite and cyclic in
the timelike coordinate.

\section{The extreme case}
In the extreme case $M=M_0$ the metric (\ref{NCSS}) becomes
\begin{equation}\label{NCSSE}
ds^2=(r-r_{0})^2\phi(r)dt^2
-\frac{dr^2}{(r-r_{0})^2\phi(r)}-r^2d\vartheta^2-r^2\sin^2{\vartheta}~d\varphi^2
\end{equation}
where $\phi$ is a differentiable and not vanishing function in the interval $[0,\infty)$.
In order to derive the maximal extension we shall follow the procedure adopted in \cite{Carter}.
To this purpose we consider the surface $\{\vartheta=\mbox{const},~\varphi=\mbox{const}\}$ and
we write (\ref{NCSSE}) as follows:
\begin{equation}\label{NCSSER}
ds^2=(r-r_{0})^2\phi(r)\left[dt
-\frac{dr}{(r-r_{0})^2\phi(r)}\right]\left[dt
+\frac{dr}{(r-r_{0})^2\phi(r)}\right]
\end{equation}
By introducing null coordinates $p$ and $q$ defined as
\[
p:=t+r^{*},\qquad q:=t-r^{*},
\]
\begin{equation}\label{tort}
r^{*}:=\int\frac{dr}{(r-r_{0})^2\phi(r)}=-\frac{1}{(r-r_{0})\phi(r_0)}
-\frac{\phi^{'}(r_0)}{\phi^2(r_0)}\ln{(r-r_0)}+\mathcal{O}(r-r_0)
\end{equation}
our metric becomes
\begin{equation}\label{NCSSERF}
ds^2=(r-r_{0})^2\phi(r)dp~dq-r^2d\vartheta^2-r^2\sin^2{\vartheta}~d\varphi^2
\end{equation}
where now $r$ is a function of $p$ and $q$. It is worth mentioning that the
surface $\{r=r_{0},~\vartheta=\mbox{const},~\varphi=\mbox{const}\}$ is made of radial null
geodesics corresponding to lines parallel to $p=\mbox{const}$ and $q=\mbox{const}$ in the $(p,q)$-plane.

\begin{figure}
\begin{center}
\scalebox{1}{\includegraphics{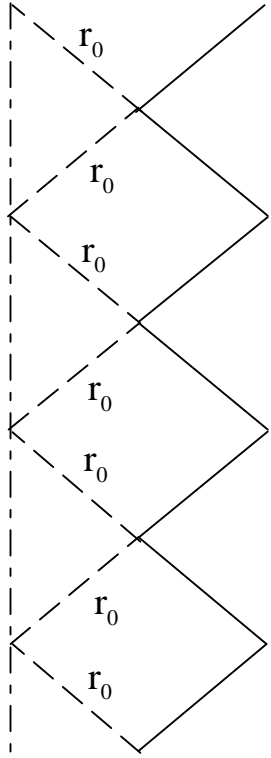}}
\caption{The conformal diagram of the maximal extension of the extreme noncommutative
inspired Schwarzschild spacetime where thin straight segments are the images of the null
infinities as $r\to\infty$, dashed segments denote the event horizon at $r_{0}$ and hatched
segments represent the deSitter core. As in the non extreme case we can identify the square
at the bottom with the next one up.}
\end{center}
\end{figure}
\label{myfigure1}

The metric (\ref{NCSSERF}) is regular for all real values of $p$ and $q$.
In order to understand where the coordinate singularity $r_0$ lies in the $(p,q)$-plane
we need to study the signs of $\phi(r_0)$ and $\phi^{'}(r_0)$. First of all, notice that
\[
\phi(r_0)=\frac{1}{2}g_{00}^{''}(r_0),\quad \phi^{'}(r_0)=\frac{1}{6}g_{00}^{'''}(r_0)
\]
Thus, the problem reduces to finding the signs of $g_{00}^{''}(r_0)$ and $g_{00}^{'''}(r_0)$.
Using the software Maple we find the following numerical values $g_{00}^{''}(r_0)\approx 0.287$
and $g_{00}^{'''}(r_0)\approx-0.277$. This implies that the coordinate singularity $r_0$ is
at $p=-\infty$ and $q=\infty$ when we move toward it from $r>r_0$ and at $p=\infty$
and $q=-\infty$ when we approach it from $r<r_0$, keeping in mind that these two regions will
be covered by two different coordinate patches. With respect to the first coordinate patch
the spatial infinity is at $p=\infty$ and $q=-\infty$. Again we can make the infinities finite
by means of the transformation $p=\tan{P}$ and $q=\tan{Q}$, the nice feature of which is that
we can lay the images of $r=r_0$ side by side in the two patches. With respect to
the coordinate patch in the region $r>r_0$ the image of $r=r_0$ is the point
with coordinates $P=-\pi/2$ and $Q=\pi/2$ whereas the choice of the coordinate
patch relative to the region $r<r_0$ sends $r_0$ to $P=\pi/2$ and $Q=-\pi/2$. in passing we note
that spacelike infinity is mapped also to the point $(\pi/2,-\pi/2)$ in the $(P,Q)$-plane.
Putting all this information together we can construct an infinite chain of conformal diagrams as
shown in Fig.2. In the present case the spacelike variable $r$
does not become timelike when we cross the event horizon. As we did
in the non extreme case with respect to the thick-line rectangle
we can identify the square at the bottom with the square at the top.
Proceeding like that we obtain a manifold which is cyclic in the timelike coordinate.

\section{Conclusions}
If we divide the quantum black hole physics into the three
categories: (i) problems zeroing around the central singularity
(final stage of a black hole) \cite{Avoidence, Adler1}, (ii) those
zeroing around the horizon (Hawking evaporation) \cite{Hawking}
and (iii) those around the question regarding the existence of
bound orbits in the outer regions \cite{Rest3}, this article
addresses the first issue in the framework of noncommutative
geometry. We found a maximal singularity-free extension of the
noncommutative geometry inspired Schwarzschild metric. The new
coordinate chart we derived in the text has the advantage that we
can illustrate more clearly the overall topology of the non
extreme and extreme noncommutative geometry inspired Schwarzschild
manifold. The Penrose diagrams are shown in Figs.$1$ and $2$,
respectively. The most striking feature of the manifold is black
hole tunnels connecting different universes and/or a cyclic
structure of the manifold in the time coordinate after
identification of two parts of the Penrose diagram (see Fig. 1) is
performed. This together with the absence of a central singularity
reveals the main difference as compared to the classical
Schwarzschild black hole structure. By some minor modification the
method we used can be applied to derive the maximal extension of
the noncommutative geometry inspired Reissner-Nordstr\"{o}m black
hole \cite{Nico3}. Finally, the stability issue of the tunnels has
been examined in \cite{miPiero}, where it has been shown that the
Cauchy horizon of the noncommutative geometry inspired
Schwarzschild black hole is stable under massless scalar
perturbations governed by a wave equation modified accordingly to
noncommutative geometry.

\end{document}